\documentclass[a4paper,final,12pt]{article}
\usepackage{graphicx}
\begin{document}
\newcommand{\gsim}{\buildrel>\over{_\sim}}
\newcommand{\lsim}{\buildrel<\over{_\sim}}
\renewcommand{\thefootnote}{\fnsymbol{footnote}}
\newcommand{\tchi}{\tilde{\chi}}
\newcommand{\psla}{p\kern-.45em/}
\newcommand{\esla}{E\kern-.45em/}
\setcounter{page}{0}
\thispagestyle{empty}
\begin{flushright}
YITP--99--4 \\
TU--559 \\
January 1999 \\
\end{flushright}

\vspace{2cm}

\begin{center}
{\large\bf Neutralino decays at the CERN LHC}
\end{center}
\baselineskip=32pt

\centerline{Mihoko M. Nojiri$^a$ and Youichi Yamada$^b$}

\baselineskip=22pt

\begin{center}
\footnotesize\it
$^a$\, YITP, Kyoto University, Kyoto 606-8502, Japan \\

$^b$\,Department of Physics, Tohoku University, Sendai 980-8578, Japan
\end{center}

\vspace{1cm}

\begin{abstract}
We study the distribution of lepton pairs from 
the second lightest neutralino decay 
$\tchi^0_2\rightarrow\tchi^0_1 l^+l^-$. 
This decay mode is important to 
measure the mass difference between $\tchi^0_2$ and the 
lightest neutralino $\tchi^0_1$, which helps to 
determine the parameters of 
the minimal supersymmetric standard model at the CERN LHC. 
We found that the decay distribution strongly depends on 
the values of underlying MSSM parameters.
For some extreme cases, the amplitude near the end point 
of the lepton invariant mass distribution can 
be suppressed so strongly that one 
needs the information of the whole $m_{ll}$ 
distribution to extract $m_{\tchi^0_2}-m_{\tchi^0_1}$.    
On the other hand, if systematic errors on the acceptance can be 
controlled, this distribution can be used to constrain 
slepton masses and the $Z\tchi^0_2\tchi^0_1$ coupling. 
Measurements of the velocity distribution of $\tchi^0_2$ 
from samples near the end point of the $m_{ll}$ distribution,
and of the asymmetry of the $p_T$ of leptons, would be useful 
to reduce the systematic errors. 

\end{abstract}

\vspace{2cm}

\vfill

\pagebreak

\normalsize\baselineskip=15pt
\setcounter{footnote}{0}
\renewcommand{\thefootnote}{\arabic{footnote}}

\section{Introduction}
If supersymmetry is  realized in nature, it promises 
exciting possibilities for future collider physics
--- the discovery of sparticles. If 
the scale of supersymmetry breaking is 
around 1 TeV (as preferred by fine-tuning arguments related to
problems in the Higgs sector), many 
sparticles will be produced at future colliders 
such as the Fermilab Tevatron upgrade, the 
CERN Large Hadron Collider (LHC), or future 
$e^+e^-$ colliders proposed by DESY, KEK, and SLAC. 
Sparticles produce unique signatures in the detectors and will 
be discovered quite easily. 

It has been pointed out that one cannot only discover those 
sparticles, but can also study their detailed nature in 
future $e^+e^-$ colliders. 
By measuring sparticle masses, 
production cross sections for a
polarized electron beam, and other distributions, 
we will measure soft supersymmetry breaking mass parameters 
\cite{JLC1,tsuka,FPMT,BMT,NFT}
and prove supersymmetric relations \cite{susy}.
Studies at future $e^+e^-$ colliders should reveal 
details of the mechanism to break supersymmetry.

Corresponding studies for the LHC have been performed
in \cite{snowmass,hin1} in the framework 
of the minimal supergravity model.
These analyses show that a precise determination of the 
model parameters is possible; the LHC is especially powerful
when the gluino decay  $\tilde{g}\rightarrow b\tilde{b}$ 
followed by $\tilde{b}\rightarrow b\tchi^0_2$, 
$\tchi^0_2\rightarrow l^+ l^- \tchi^0_1$ can be 
identified by tagging the bottom jets and the hard leptons. 
One of the key tricks of the studies 
is the measurement of the end point of the invariant mass distribution 
of the lepton pair with same flavor and opposite charges. 
The end point determines the mass difference between the second 
lightest and the lightest neutralino \cite{BHT},
$m_{\tchi^0_2}-m_{\tchi^0_1}$. The kinematically 
constrained nature of the end point samples allows one 
to reconstruct the decay chains to determine 
the parameters of minimal supergravity model \cite{snowmass,hin1}.  
At the end, all or some of the parameters in the model
would be determined.\footnote{A similar 
analysis can be done \cite{hin2} for the
parameter dependence of gauge mediated 
supersymmetry breaking models \cite{GMSB}.} 

However, the minimal supergravity model is not the only 
model for spontaneous supersymmetry breaking 
in a phenomenologically consistent manner. 
Moreover, there are various 
possibilities even within the supergravity model, 
as summarized in Ref. \cite{snowmassth} for example. 
Other mechanisms to break supersymmetry (SUSY) are also 
discussed extensively \cite{GMSB,DP}.

Assuming that the LHC can determine 
all parameters within the minimal supergravity model by looking 
into some signal distributions, the next question 
is what we should look into to over-constrain 
the model, or to determine deviations from the minimal 
supergravity model. As we mentioned before, studies in 
this direction  have been done in great detail for future  
$e^+e^-$ colliders. For hadron colliders, such a study would 
be complicated because %of the large QCD backgrounds, 
%and the tight cuts needed to reduce them. 
more than one SUSY channel can contribute to a particular 
class of events since different sparticles are simultaneously
produced. 

In this paper, we discuss the parameter dependence 
of the distribution of the three body decay 
$\tchi^0_2\rightarrow \tchi^0_1 l^+l^-$. 
The decay branching 
ratio depends on parameters in the minimal supersymmetric 
standard model (MSSM) rather sensitively due to
(possibly negative) interference between slepton
and $Z^0$ exchange contributions \cite{BT,BCKT,MKKW,AM}. 
We point out that not only the branching ratio 
but also the decay distribution is sensitive to the 
MSSM parameters, 
giving us an extra handle to determine 
$m_{\tilde{l}}$, $\tan\beta$, $M_i$ and $\mu$, 
independent of the SUSY breaking mechanism. 

The organization 
of this paper is as follows. In section 2, 
we discuss the MSSM parameter dependence of the
lepton invariant mass distribution arising 
from $\tchi^0_2$ decay. Near the region where 
interference between $Z^0$ exchange and 
$\tilde{l}$ exchange becomes important, the 
decay distribution is very sensitive to
$m_{\tilde{l}}$. We point out 
that studying the whole distribution of the lepton invariant 
mass $m_{ll}$ can be very
important even for extracting the end point; 
events near the end point 
could be so few that one could misidentify it without 
this information. 
We also discuss the chiral structure of the 
amplitude, and show an interesting 
parameter dependence of the tau polarization
in $\tchi^0_2\rightarrow$ $\tau^+\tau^-\tchi^0_1$ decays. 

In section 3, we discuss effects of cuts 
on observed distributions. Though $m_{ll}$ of lepton pairs
from $\tchi^0_2$ decay is independent of the $\tchi^0_2$ boost, 
each lepton energy depends on the parent $\tchi^0_2$ 
momentum. Therefore we expect two apparent 
sources of systematic error: (A) uncertainty of the lepton  energy 
distribution in the $\tchi^0_2$ rest frame, and (B) uncertainty of 
the parent $\tchi^0_2$ momentum distribution.
It would be helpful to reduce these two errors 
in order to maximize the physical information 
that can be extracted from the observed $m_{ll}$ 
distribution. We argue that a measurement of the asymmetry of lepton 
energies would reduce the systematic errors from (A), 
while the lepton energy distribution near the end point of the $m_{ll}$ 
distribution constrains (B).
In section 4 we study if the $m_{ll}$ 
distribution constrains $m_{\tilde{l}}$ 
and the parameters in the neutralino mass matrix. Section 5 is 
devoted to discussion and comments.

\section {Lepton Invariant Mass Distribution}

The branching ratio of the leptonic decay of the second lightest 
neutralino, $\tchi^0_2\rightarrow\tchi^0_1l^+l^-$, is known to be 
very sensitive to the values of the underlying parameters. 
Three body decays of $\tchi^0_2$ are dominant \cite{BT,BCKT,MKKW} 
as long as $\tchi^0_2\rightarrow\tchi^0_1Z^0$, 
$\tchi^0_2\rightarrow\tilde{l}l$ are not open. 
The dependence is enhanced by the negative interference between the
decay amplitude from $Z^0$ exchange and that from
slepton exchange. In the minimal supergravity 
model, the interference would be maximal
when slepton masses are around 200 GeV \cite{MKKW}. 

In this section we show that the effect of the interference 
appears not only in the branching ratios, but also 
in the decay distributions, such as the distribution of the 
invariant mass $m_{ll}$ of two the leptons. 

We first show the differential partial width of the decay 
$\tchi^0_A\rightarrow\tchi^0_B\bar{f}f$ 
for a general light fermion $f$. We assume that the mass and Yukawa 
coupling of $f$, 
and left-right mixing of $\tilde{f}$ are negligible. 
The decay amplitude then consists of the $Z^0$ channel and the 
$\tilde{f}_{L,R}$ channel. 
The squared, spin averaged\footnote{The polarization of $\tchi^0_2$ 
might affect observed decay distributions. The effect depend on 
the process for $\tchi^0_2$ production, and is not discussed 
in this paper.} amplitude $\vert{\cal M}\vert^2$ of the decay 
$\tchi^0_A(p)\rightarrow\tchi^0_B(\bar{p})\bar{f}(\bar{q})f(q)$ 
is written as follows:
\begin{eqnarray}\label{e1}
\vert{\cal M}\vert^2 &=& 
2 (A_{LL}^2+A_{RR}^2) (1-y) (y-r_{\tilde{\chi}_B}^2) 
+2 (A_{LR}^2+A_{RL}^2) (1-x) (x-r_{\tilde{\chi}_B}^2) 
\nonumber\\
&&-4 (A_{LL}A_{RL}+A_{RR}A_{LR}) r_{\tilde{\chi}_B} z,
\end{eqnarray}
with
\begin{eqnarray}
A_{LL} &=&  \frac12 g_Z^2 \frac{z^{(\tilde \chi^0)}_{BA} z^{(f)}_L}{z-r_Z^2} 
           -\sum_X \frac12 g_2^2 \frac{a^f_{AX} a^f_{BX}}
                              {y-r_{\tilde{f}_X}^2},
\nonumber\\
A_{RL} &=& -A_{LL}(y\leftrightarrow x), \nonumber\\
\nonumber\\
A_{LR} &=&  \frac12 g_Z^2 \frac{z^{(\tilde \chi^0)}_{BA} z^{(f)}_R}{z-r_Z^2} 
           +\sum_X \frac12 g_2^2 \frac{b^f_{AX} b^f_{BX}}
                              {x-r_{\tilde{f}_X}^2},
\nonumber\\
A_{RR} &=& -A_{LR}(y\leftrightarrow x), \label{e2}
\end{eqnarray}
where $z^{(f)}_L= T_{3fL}-Q_f s_W^2$, $z^{(f)}_R=-Q_f s_W^2$, with 
$s_W^2=\sin^2\theta_W$. 
The forms of the $\tchi^0\tchi^0Z^0$ couplings $z^{(\tchi^0)}_{BA}$ and 
$\tchi^0f\tilde{f}$ couplings ($a^f_{AX}$, $b^f_{AX}$) are 
given in Appendix. 
$(x,y,z)$ are phase space parameters of the decay defined as 
\begin{equation}\label{e3}
x=(\bar{p}+ q)^2/m^2_{\tchi^0_A},\;\;
y=(\bar{p}+ \bar{q})^2/m^2_{\tchi^0_A},\;\;
z=(q+ \bar{q})^2/m^2_{\tchi^0_A},
\end{equation}
with $x+y+z=1+m_{\tchi^0_B}^2/m_{\tchi^0_A}^2$.
For convenience, we have introduced mass ratios 
$r_{\tilde{\chi}_B}$, $r_Z$, and $r_{\tilde{f}_X}$ as 
\begin{equation}\label{e4}
r_{\tchi_B} = m_{\tchi^0_B}/m_{\tchi^0_A},
\;\;
r_Z = m_Z/m_{\tchi^0_A},
\;\;
r_{\tilde{f}_X} = m_{\tilde{f}_X}/m_{\tchi^0_A}.
\end{equation}
Note that $(A_{LL},A_{RL})$ involve $(f_L,\bar{f}_R)$ while 
$(A_{LR},A_{RR})$ involve $(f_R,\bar{f}_L)$.
They do not interfere for $m_f=0$.

The partial decay width is given by 
\begin{equation}\label{e5}
\frac{d\Gamma}{dx\,dy}(\tchi^0_A\rightarrow\tchi^0_B\bar{f}f)=
\frac{N_C}{256 \pi^3} m_{\tchi^0_A} \vert{\cal M}\vert^2(x,y,
z=1+ r_{\tchi_B}^2-x-y),
\end{equation}
where $N_C=3(1)$ for $f=q(l)$. 
The range of ($x$, $y$) is given by the conditions
\begin{eqnarray}\label{e6}
&z (xy-r_{\tilde{\chi}_B}^2) \ge 0,&
\nonumber\\
&r_{\tchi_B}^2 \le x \le 1,&
\nonumber\\
&r_{\tchi_B}^2 \le y \le 1,&
\nonumber\\
&x+y+z=1+ r_{\tchi_B}^2.&
\label{psd}
\end{eqnarray}

Now we consider the case of $\tchi^0_2$ decay into $\tchi^0_1l^+l^-$ 
under the assumption that all two body decays of $\tchi^0_2$ 
are kinematically forbidden.\footnote{In some region of parameter 
space, we may study the three body decays of $\tchi^0_2$ even 
if two body decays are open \cite{IK}. Such a study 
is beyond the scope of this paper. } 

In the phase space of the decay 
$\tchi^0_2\rightarrow\tchi^0_1l^+l^-$, 
the $Z^0$ exchange amplitude and $\tilde{l}$ 
exchange amplitude behave differently. 
When the $Z^0$ contribution dominates, 
distributions are enhanced in the region of 
large $m^2_{ll}=zm_{\tchi^0_2}^2$.
In contrast, when the $\tilde{l}$ exchange contribution 
dominates, distributions are enhanced in regions with 
large $x$ and/or large $y$, therefore 
in small $m_{ll}$ and large 
$\vert E^{\rm rest}_{l^-}-E^{\rm rest}_{l^+}\vert$ region. 
Here 
\begin{equation}\label{e7}
E^{\rm rest}_{l^+}=(1-x){m_{\tchi^0_2}}/2,\;\;
E^{\rm rest}_{l^-}=(1-y){m_{\tchi^0_2}}/2,
\end{equation}
are lepton energies in the $\tchi^0_2$ rest frame. 

We consider the case where $2M_1\sim M_2 \ll |\mu|$, a typical case 
in the minimal supergravity model. In this case, $\tchi^0_2$ is 
Wino-like and $\tchi^0_1$ is Bino-like. 
An interesting property in this case is that the $Z^0$ and $\tilde{l}$ 
amplitudes could be of comparable size in some region of 
phase space. Furthermore, their interference is generally 
destructive for leptonic decays. These effects cause 
complicated situations, which we discuss below. 

We show numerical results for the $m_{ll}$ distribution. 
For illustration, 
we use two sets of parameters for the neutralino sector, (A) and (B), 
shown in Table \ref{nsets}. 
\begin{table}[htb]
\begin{center}
\begin{tabular}{|c|c|c|c|c|} \hline
$\Biggl.$set & $M_1$ & $M_2$ & $\mu$ & $\tan\beta$ \\[2mm]\hline
$\Biggl.$(A)   & 70    & 140   & --300 & 4           \\[1mm]\hline
$\Biggl.$(B)   & 77.6  & 165   & 286   & 4           \\[1mm]\hline
\end{tabular}
\end{center}
\caption{\footnotesize Parameter sets for neutralinos. 
All entries with mass units are in GeV.}
\label{nsets}
\end{table}
These values are fixed to give the same masses for three inos, 
($m_{\tchi^0_1}$, $m_{\tchi^0_2}$, $m_{\tchi^+_2}$) $=$ 
(71.4, 140.1, 320.6) GeV. For calculating the branching ratios, 
we take generation-independent slepton masses and 
a universal soft SUSY breaking squark mass $m_{\tilde{Q}}=500$ GeV.

\begin{figure}[htbp]
\begin{center}
\includegraphics[width=6cm,angle=-90]{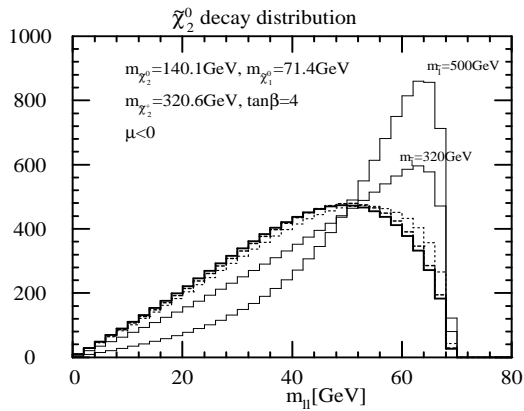}
\end{center}
\caption{\footnotesize  Invariant mass distribution of the lepton pairs  
from $\tchi^0_2$ three body decay. The neutralino parameters 
are taken as set (A) (see Table \ref{nsets}), and universal 
slepton masses $m_{\tilde{l}}=m_{\tilde{l}_L}=m_{\tilde{l}_R}$ 
are 170 GeV (thick solid), 220 GeV 
(dashed), 270 GeV (dotted), 320 GeV, and 500 GeV. 
For $m_{\tilde{Q}}=500$ GeV, 
$Br(\tchi^0_2 \rightarrow $ $e^+e^-\tchi^0_1)=$11\%, 9.5\%,
 4.1\%, 2.1\%, 1.9\%, respectively. 
The total number of events of each curve is  $10^4$. 
}
\label{fig1}
\end{figure}

In Fig. 1, we show the $m_{ll}$ distribution 
of the decay $\tchi^0_2\rightarrow l^+ l^- \tchi^0_1$ 
for parameter set (A) and 
varying $m_{\tilde{l}}$ from 170 GeV to 500 GeV. 
\footnote{Heavy sleptons and 
Bino-like $\tchi^0_1$ is cosmologically disfavored because it leads to a
large relic mass density \cite{cosm}.
However, this constraint can easily be evaded 
if $\tchi^0_1$ can decay, or if there is late time 
entropy production.
}

Because $m_{\tchi^0_2}$ and $m_{\tchi^0_1}$ are fixed, the 
end points of the distributions $m_{ll}^{\rm max}=68.7$ GeV 
are same for each curve, while the shape of the distribution changes 
drastically with slepton mass. 
For a slepton mass of 170 GeV (thick solid line), 
the decay proceeds dominantly through slepton 
exchanges, therefore the $m_{ll}$ distribution is suppressed 
near $m_{ll}^{\rm max}$. On the other hand, once slepton exchange 
is suppressed by its mass, $Z^0$ exchange dominates 
and the distribution peaks sharply near $m_{ll}^{\rm max}$. 
Notice that
the leptonic  branching ratio decreases 
as $m_{\tilde{l}}$ increases, but remains larger than $\sim$2\% 
for the values of $m_{\tilde{l}}$ taken in the figure.

\begin{figure}[htbp]
\begin{center}
\includegraphics[width=7cm,angle=-90]{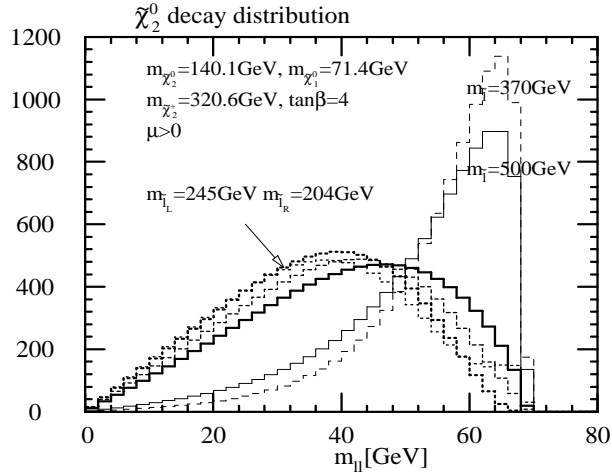}
\end{center}
\caption{\footnotesize $m_{ll}$ distribution for parameter 
set (B) ($\mu>0$). $m_{\tilde{l}_L}=m_{\tilde{l}_R}$ are  
170 GeV (thick solid), 220 GeV (dashed), 250 GeV (dotted), 370 GeV, 
and 500 GeV. The branching ratio 
$Br(\tchi^0_2 \rightarrow e^+e^-\tchi^0_1)$ 
for $m_{\tilde{Q}}=500$ GeV is 
6.6\%, 2.9\%, 0.9\%, 1\%, 1.8\%, respectively, 
The thick dashed line is an example with complete cancellation 
of the amplitude near the upper end point. The branching 
ratio to $e^+e^-\tchi^0_1$ is 1.8\% for this case. 
The total number of the events is $10^4$ for each curve.}
\label{fig2}
\end{figure}

In Fig. 2, we show an example for $\mu>0$, parameter set (B).
The dependence on the slepton mass is different from the previous 
case. As $m_{\tilde{l}}$ increases from 170 GeV, $m_{ll}$ 
distribution becomes softer. For 
$m_{\tilde{l}}> 250$ GeV, a second peak appears 
due to strong cancellation of $Z^0$ exchange 
and slepton exchange contributions for a certain value of 
$m_{ll}$.  At the same time, the branching ratio reaches its minimum 
at $m_{\tilde l}\sim 300$ GeV, much less than 1\%. 
For $m_{\tilde{l}}\gg 370$ GeV, it increases again 
above 1\%. 

Notably, one can find slepton masses 
where a complete cancellation occurs very 
close to the end point $m^{\rm max}_{ll}$ of the $m_{ll}$ distribution. 
The thick dashed line shows distribution for 
$m_{\tilde{l}_L}=245$ GeV and $m_{\tilde{l}_R}=204$ GeV. 
Events near the end point 
($m^{\rm max}_{ll}-m_{ll}< 4$ GeV) becomes too few, 
and it is very hard to observe the real end point 
for this case.

The lepton invariant mass distribution is an important tool for 
studying supersymmetric models at hadron colliders. 
In \cite{hin1}, a case study is done in a scenario where 
decays $\tilde{g}\rightarrow b\tilde{b}$
and $\tilde{b}\rightarrow b\tchi^0_2$ 
have substantial branching ratios. 
$\tchi^0_2$ production is enhanced by the
large gluino production cross section, and 
the $S/N$ ratio could be improved 
substantially by requiring three or four bottom 
jets in the final states. For 
$Br(\tchi^0_2\rightarrow$ 
$e^+e^- \tchi^0_1)\sim 16$ \%, 
$S/N$ goes well above 10. 
Even if gluino decay into $\tilde{b}b$ is closed, 
the lepton invariant mass distribution can be measured (case 4 of 
\cite{hin1}). In this case, one can subtract most
backgrounds using lepton pair samples with opposite charge and 
different flavors.

The most important aspect of the lepton invariant mass distribution 
in these studies is the determination of the 
end point, which is expected to 
coincide with $m_{\tchi^0_2}-m_{\tchi^0_1}$. Furthermore,
choosing events near the end point helps 
one to relate the velocity of the two lepton system $\beta_{ll}$ to 
that of the neutralinos $\beta_{\tchi^0_2}$ and $\beta_{\tchi^0_1}$, 
so that one can reconstruct the cascade decay chain. 
Errors on the mass difference of 2\% $\sim$ 0.1\%
are claimed depending on statistics \cite{hin1}.

The slepton mass dependence of $\tchi^0_2$ decay distributions
suggests that not only the end point 
of the distributions but also the distributions themselves 
contain information about the underlying parameters
such as $m_{\tilde{l}}$. 
The negative side of this is that the fitted
end point may depend on the assumed values of these
parameters, introducing
additional systematic errors in the fit. 
For some extreme case shown in Fig. 2, the observed end point of 
the lepton invariant mass distribution does {\it not} coincide with 
$m_{\tchi^0_2}-m_{\tchi^0_1}$. 
Note that realistic simulations including 
the parameter dependence of the decay distribution
are not available for hadron colliders so far. 
In the commonly used Monte Carlo (MC) simulators 
ISAJET \cite{ISAJET} and SPYTHIA \cite{SPYTHIA}, the three body decay 
distribution of sparticles is approximated by 
the phase space distribution, while branching ratios 
are calculated by full expressions.\footnote{The most recent ISAJET 
release (ISAJET 7.43) allows to simulate 
the effect of exact matrix elements for all three body decay 
distributions. 
In the codes used for CERN LEP \cite{SUSYGEN} and JLC1 \cite{JLCcode} 
studies, the effect is already included.}
See section 3 (especially Fig. 4) for comparison between 
real $m_{ll}$ distribution and the 
distribution in phase space approximation.

For the case shown in Figs.~1 and 2, 
$Br(\tchi^0_2\rightarrow\tchi^0_1 l^+l^- )$ is not always large. 
It has not been yet studied systematically 
if it is possible to measure the lepton decay distribution 
at future hadron and $e^+e^-$ colliders when 
$Br(\tchi^0_2\rightarrow\tchi^0_1 l^+l^- )$ is small. 
Given small Standard Model backgrounds, 
a study of distribution may be  possible 
when $Br(\tchi^0_2\rightarrow$ 
$e^+e^- \tchi^0_1)\gsim2$\% for not too heavy gluino.
At future lepton colliders, it is very easy to get 
a clean signal. However, the production 
cross section is rather small for heavy sleptons. 
At $e^+e^-$ collider 
with large luminosity ($\int {\cal L} > 500$ fb$^{-1}$/yr) 
\cite{DESY}, the region with 
$\sigma(e^+e^-\rightarrow\tchi^0_2\tchi^0_2)>20$ fb 
may be studied with samples containing more than 1000 events, 
even if $Br(\tchi^0_2\rightarrow
e^+e^- \tchi^0_1)\sim$ a few \%. 

In Figs. 1 and 2, we found that the decay 
distributions depend on sign($\mu$). 
The difference comes from  neutralino mixings. For the 
parameter set (A), $\tchi^0_2$ has small Bino component; 
$(N_{21},N_{22},$
$N_{23},N_{24}) =$ 
$(7.25\times 10^{-4},0.957,-0.281,$
$-6.90\times 10^{-2})$
where $\tchi^0_i=N_{i1}\tilde{B}+N_{i2}\tilde{W}^3
+N_{i3}\tilde{H}^0_1+N_{i4}\tilde{H}^0_2$. 
It does not couple to $\tilde{l}_R$ effectively [see Eq.~(\ref{e13})], 
therefore decay into $l_R$ proceeds 
dominantly through $Z^0$ exchange.
In Fig. 3(a) we show the $m_{ll}$ distribution for 
$l^+l^-_R$(dotted lines)  and $l^+l^-_L$(dashed lines)
for $m_{\tilde{l}}=220$ GeV(thick lines) and 270 GeV(thin lines). 
When $m_{\tilde{l}}$ increases from 220 GeV 
to 270 GeV, 
$\vert {\cal M}_L\vert^2=\vert {\cal M}(\tchi^0_1l^+l^-_L)\vert^2$
is suppressed near $m_{ll}^{\rm max}$ and also becomes smaller 
due to interference. 
The relative importance of  $\vert{\cal M}_R\vert$ 
increases, leading to a total distribution that is more strongly peaked 
near $m^{\rm max}_{ll}$. 
On the other hand, for parameter set (B) ($\mu>0$), $\tchi^0_2$ 
still has substantial Bino component; 
$(N_{21},N_{22},$
$N_{23},N_{24})$ 
$=( 0.196,0.902,-0.322,0.212)$.
As $m_{\tilde{l}}$ increases, 
negative interference appears 
near $m_{ll}= m^{\rm max}_{ll}$ for both 
$\vert{\cal M}_L \vert$ and $\vert{\cal M}_R\vert$,
leading to a suppression of the $m_{ll}$ distribution near $m_{ll}^{\rm max}$. 
  
\begin{figure}[htbp]
\begin{center}
\includegraphics[width=5.5cm,angle=-90]{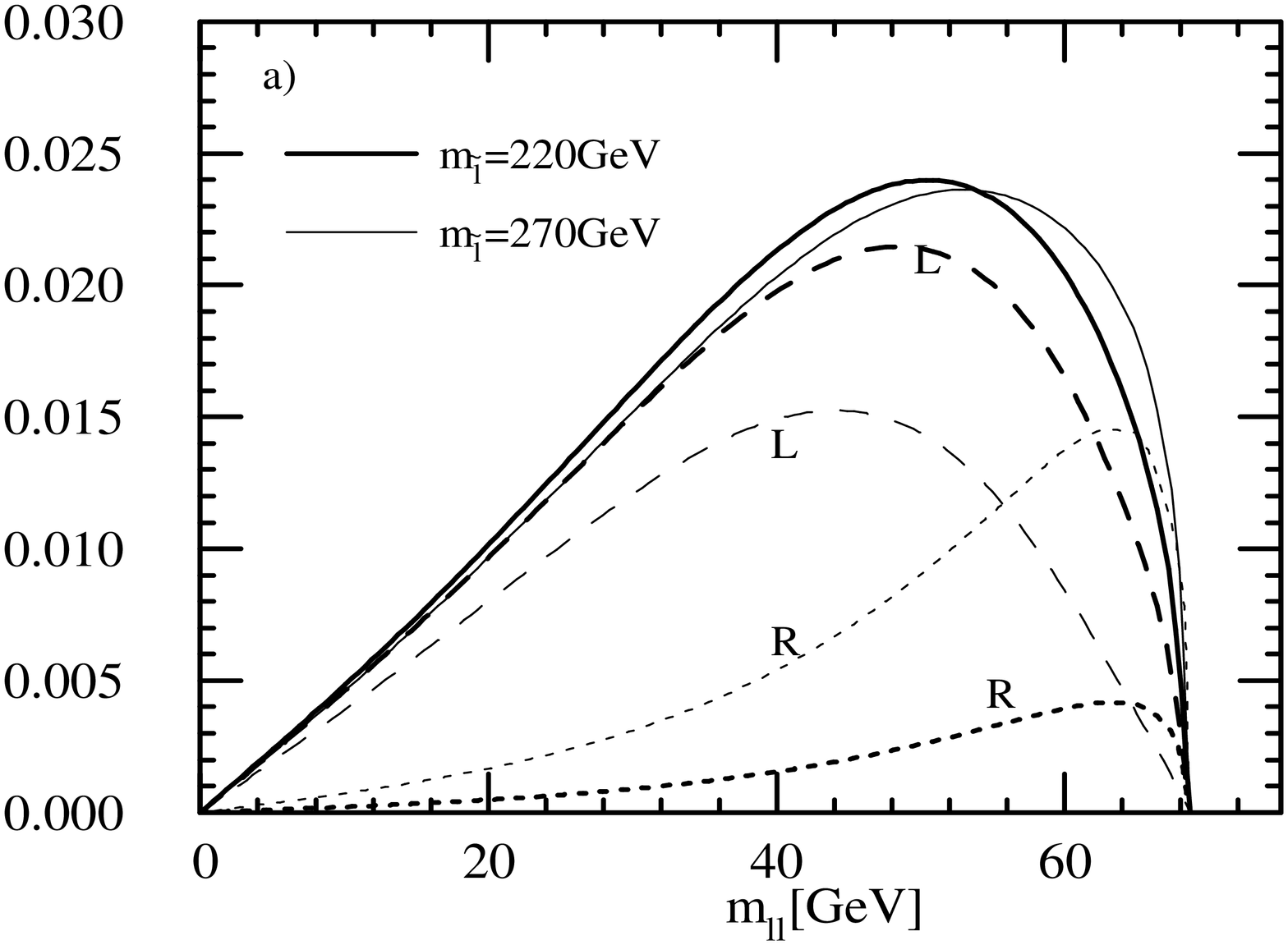}
\hskip 2cm 
\includegraphics[width=5.5cm,angle=-90]{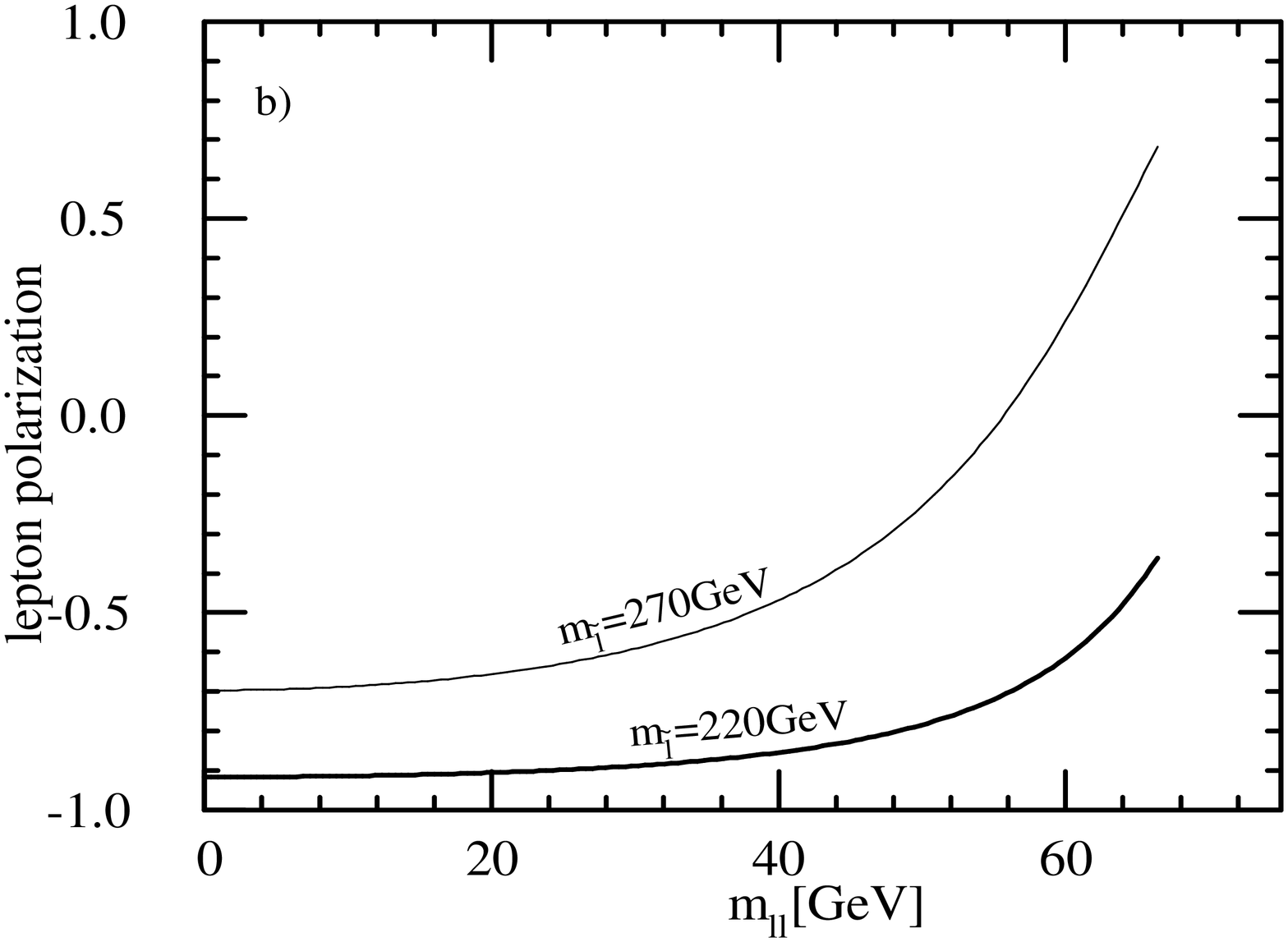}
\end{center}
\caption{
\footnotesize (a) $m_{ll}$ distribution from 
$\tchi^0_2\rightarrow \tchi^0_1 l^+l^-$ 
decays for $m_{\tilde{l}}=220$ GeV and 
$m_{\tilde{l}}=270$ GeV and parameter set (A).  
The dashed lines show $l^+l^-_L$ distribution, 
the dotted lines for  $l^+l^-_R$ distribution, the solid lines 
for total. The curves are normalized so that the 
total widths are the same.
(b) Polarization of lepton $l^-$ 
[$\equiv (N_R-N_L)/(N_R+N_L)$] arising from $\tchi^0_2$ decay. 
Average polarization is $P_{ave}= -0.839$ for $m_{\tilde{l}}=220$ GeV 
and $-0.428$ for $m_{\tilde{l}}=270$ GeV. 
}
\label{fig3}
\end{figure}

In Fig. 3(a), negative interference reduces the amplitude near 
$m^{\rm max}_{ll}$ for $l^-_L$, while the amplitude for 
$l^-_R$ is increased. This causes a strong $m_{ll}$ dependence 
of the lepton polarization $P_l$. 
In Fig. 3(b), we plot this polarization 
as a function of $m_{ll}$ for different slepton masses.  
They differ dramatically near $m^{\rm max}_{ll}$. 
In neutralino decays into $\tau^+\tau^-\tchi^0_1$, 
$P_{\tau}$ may be observed through 
the decay distributions of the $\tau$ leptons.
In $\tau^{\pm}\rightarrow\rho^{\pm}$ $\rightarrow\pi^{\pm}\pi^0$, 
$a_1\rightarrow\pi^{\pm}\pi^0\pi^0$ decays, 
the $E_{\pi^{\pm}}/E_{jet}$ distribution depends on the parent 
$\tau$ polarization drastically \cite{tau}. 
Experimentally, the fine momentum resolution \cite{JLC1,snowmass} of 
the detector could be used to identify these tau decay products \cite{NFT}
at future $e^+e^-$ colliders. At hadron colliders, the momentum of 
charged tracks and the energy deposited in the
electromagnetic calorimeter would give the same information. 
Implications of $\tau$ polarization measurements in analyses of 
new physics have been discussed for charged Higgs bosons produced at the 
upgraded Tevatron and LHC colliders \cite{tauLHC}, 
and for $\tilde{\tau}$ at $e^+e^-$ colliders \cite{nojiri,NFT}.

Due to the missing tau neutrinos, one would not be able to measure 
the invariant mass of two tau leptons. Nevertheless the 
$P_{\tau}$ dependence on the invariant mass of the two $\tau$ jets
might be seen in future collider experiments.
Note that in $\tau\rightarrow \rho$ or $ a_1$ decays, the final 
vector meson carries a substantial part of the parent $\tau$ momentum, 
therefore the smearing of the distribution is less 
severe
than for decays into $\pi^{\pm}$, $\mu$, and $e$. 

Several comments are in order. 
In Eq. (\ref{e1}), we have neglected the Yukawa couplings 
of leptons and slepton left right mixing. 
For $\tilde{\tau}$, these effects could be very important 
if $\tan\beta$ is large. 
Notice that their leading contribution flips the chirality 
of the $\tau$ lepton \cite{nojiri}. For three body decays, 
studying the correlation 
of two tau decay distributions would reveal the helicity 
flipping and conserving contributions separately. 

In most numerical calculations in this paper, we assume 
universality of slepton masses. However, 
in supersymmetric model, staus could be lighter 
than the other sleptons for various reasons. 
The running of stau soft SUSY breaking masses from the Planck 
scale \cite{BH} and stau left-right mixing \cite{DN1} could enhance 
decays into $\tau\tilde{\tau}$ or $\tau^+\tau^-\tchi^0_1$.
Experimental consequences of such scenarios have recently been widely 
discussed \cite{neut_taua, neut_taub}.
Also models with lighter third generation sparticles
have been constructed to naturally avoid 
the flavor changing neutral current problem \cite{DP}.   
The two body decay branching ratio into $\tau\tilde{\tau}$  
or the three body decay 
branching ratio into $\tau\tau\tchi^0_1$ and the decay distribution 
might be different from those for leptons 
in the first two generations. 
The study of the $\tchi^0_2\rightarrow$ $ \tau^+\tau^-\tchi^0_1$ decay 
in addition to the other leptonic modes could be 
an important handle to identify such models.

\section{ Correcting Acceptance Errors }

In collider experiments, we need cuts to reduce 
backgrounds, and they substantially change 
distributions we are interested in. 
This effect can be corrected by Monte Carlo simulations
once model parameters are fixed. But it is still 
worthwhile to investigate how the distributions are modified 
by the cuts, and if this effect can be estimated 
in a model-independent way. 

Let us consider $\tchi^0_2$ decays at the LHC, 
where $\tchi^0_2$ comes from 
the cascade decay of heavier squarks and gluinos: 
\begin{eqnarray}\label{e8}
\tilde{q}&\rightarrow & \left\{ 
\begin{array}{l}
\tilde{g} q \ ({\rm if }\ m_{\tilde{q}}>m_{\tilde{g}} ) ,\\
q\tchi^0_2,\\
\end{array} \right. \nonumber\\
\tilde{g}& \rightarrow & \left\{
\begin{array}{l}
q \bar{q}\tchi^0_2,\\
q\tilde{q}\ 
({\rm if}\  m_{\tilde{q}}<m_{\tilde{g}} ), \\
\end{array} \right. 
\end{eqnarray}
and $\tchi^0_2$ may decay further into $l^+l^-\tchi^0_1$. 
These decay processes have been shown to have small backgrounds. 
In \cite{hin1}, $\tilde{g}$ decay into $b\tilde{b}$ 
followed by 
$\tilde{b}\rightarrow b\tchi^0_2$, and $\tilde{g}$ decay into 
$qq\tchi^0_2$ were studied. After lepton transverse momentum 
cuts, cuts on total transverse energy $E_T$ and total
missing transverse energy $\esla_T$, 
$b$ tagging (for the former), and subtraction of 
backgrounds estimated from the opposite charge -- different 
flavor lepton sample, the $m_{ll}$ distribution from $\tchi^0_2$
decays can be measured over a wide region of $m_{ll}$. 

The observed distributions are modified
by the cuts, and they depend on the decay 
processes through which $\tchi^0_2$ is produced.
One might worry that those cuts are strongly correlated with $m_{ll}$, 
i.e., particular regions of $m_{ll}$ 
have large (small) acceptance, therefore the 
whole distribution may become quite different from 
the original one. Indeed, the cuts on lepton energy 
should affect the $m_{ll}$ distribution directly. 
On the other hand, 
cuts on the total $E_T$ and $\esla_T$ should be less correlated 
with the momenta of the leptons. In the following, we therefore discuss 
the effect of cuts on lepton energies on the 
lepton invariant mass distribution. 

In Ref. \cite{hin1}, a lepton transverse momentum 
cut, $p^l_{T}>$ 10 GeV $\sim$15 GeV, is applied to all leptons. 
The $m_{ll}$ distribution after this 
cut should depend on the $E_l$ distributions of
$\tchi^0_2$ decays in the $\tchi^0_2$ rest frame, and also on
the $\tchi^0_2$ momentum distribution through the boost of leptons.
\begin{figure}[htbp]
\begin{center}
\includegraphics[width=7cm,angle=-90]{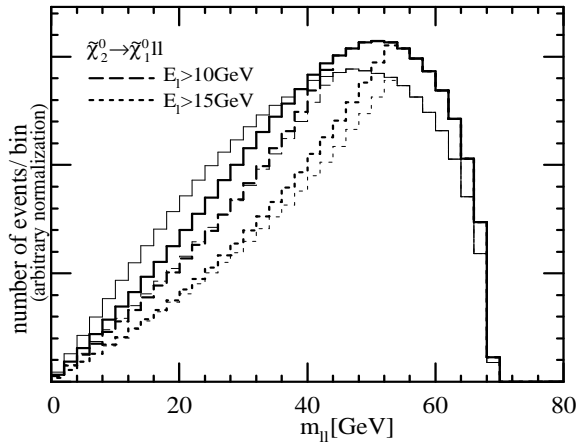}
\end{center}
\caption{\footnotesize 
Invariant mass distributions of the lepton pairs 
from $\tchi^0_2$ decay for the parameter set (A) and 
$m_{\tilde{l}}=250$ GeV (thick lines). $E_{l}>0$ (solid line),
10 GeV (dashed), and 15 GeV (dotted) is required. We also 
show the distributions in phase space approximation 
by thin lines.  
}
\label{fig4}
\end{figure}
In Fig. 4, we plot the $m_{ll}$ distribution  
for our standard parameter set
(A) and $m_{\tilde{l}}= 250$ GeV, requiring that the lepton 
energy $E_l$ in the $\tilde{\chi}^0_2$ rest frame is 
larger than 0 (thick solid), 
10 GeV (thick dashed), 15 GeV (thick dotted), respectively. 
We also plot the corresponding distributions 
in phase space approximation.

Figure 4 reproduces the quantitative effects of $p_T^l$ cut.
Because of the lepton $p_T$ cuts, leptons with 
$E_l< p_T^{l\rm cut}$ have no chance of being accepted 
unless the parent neutralino has enough transverse momentum. 
Therefore the dashed and dotted lines show quantitative 
nature of the $m_{ll}$ distribution after the $p_T^l$ cut
when the momentum of the parent neutralino 
is negligible. Typically, events with large $m_{ll}$ have 
a better chance of being accepted. 
We can see that both of the leptons have energy larger 
than 15 GeV if $m_{ll}>54$ GeV.
$E_l^{\rm rest}\sim 35$ GeV in the region close to
$m_{ll}^{\rm max}$. 
Unless leptons go in the beam
direction, they would be accepted. 
Although the actual distribution after the $p^l_{T}$ cut would 
depend on the $p_T$ distribution of $\tchi^0_2$, 
it is still qualitatively true that  events with 
small $m_{ll}$ would be more affected by the 
cuts. 
Note that, whenever a large cancellation of the amplitude 
occurs at the $m_{ll}$ end point, the $m_{ll}$ distribution 
%near the end point are also reduced substantially. 
near the end point also differs substantially from those without 
cancellation.
The acceptance near the end point should be large and should depend
on $m_{ll}$ only weakly,
so that we could distinguish a ``fake'' end point from the
kinematical one through the study of the distribution.

In Fig. 4, the $m_{ll}$ distribution 
is softer in the phase space approximation. However, when 
$E_l^{\rm rest}>10$ GeV is required, the number of
events with smaller $m_{ll}$ is reduced significantly 
compared with that for parameter set (A). The 
qualitative difference between the two curves 
becomes less significant after $E_l^{\rm rest}>10$ GeV is 
required. 

\begin{figure}[htbp]
\begin{center}
\includegraphics[width=7cm,angle=-90]{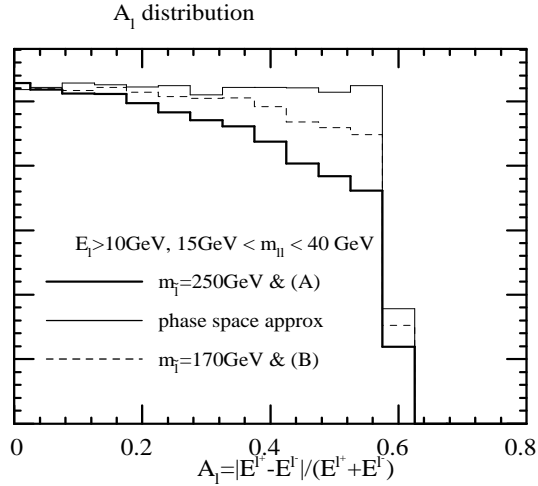}
\end{center}
\caption{\footnotesize $A_l= $ $\vert 
E^{\rm rest}_{l^+}-E^{\rm rest}_{l^-}\vert/$
$(E^{\rm rest}_{l^+}+E^{\rm rest}_{l^-})$ 
distribution of the lepton pair from $\tchi^0_2$ decays
with 15 GeV$< m_{ll}<40$ GeV and $E_l>$ 10 GeV.  
The thick solid line 
is for $m_{\tilde{l}}=250$ GeV and the parameter set (A), 
the dashed line for $m_{\tilde{l}}= 170$ GeV
and the parameter set (B), and the solid line for phase 
space approximation. 
Curves are normalized to coincide at $A_l=0$.}
\label{fig5}
\end{figure}

The difference of the acceptance in smaller $m_{ll}$ region 
can be seen when we plot lepton energy difference.
In Fig. 5 we plot $A_l=|E_{l^+}-E_{l^-}|/(E_{l^+}+E_{l^-})$ in 
the rest frame of $\tchi^0_2$ requiring $E_l>10$ GeV 
and\footnote{Here we restrict $m_{ll}$ from 15 GeV to 40 GeV. 
We may have a background of $\gamma^*\rightarrow l^+l^-$ in 
small $m_{ll}$ region. For $m_{ll}$ near the end point, 
leptons tend to have equal energy, and the contribution to 
$A_l$ is insignificant.}
$15$ GeV $ <m_{ll}<40$ GeV. The distribution reaches 
its maximum at $A_l=0$ for the parameter set (A) with
$m_{\tilde{l}}=250$ GeV (thick line), 
while for phase space approximation (solid) 
the $A_l$ distributions are roughly flat. 
This is because the amplitude near $E_l^{\rm rest}=0$ is 
suppressed, and agrees with the quantitative difference of the 
acceptance in small $m_{ll}$ region found in Fig. 4.
The acceptance should also depend on underlying MSSM parameters.
We show the distribution for the parameter set (B) and 
$m_{\tilde{l}}=170$ GeV, which has similar $m_{ll}$ 
distribution to $\mu<0$ case. 
Due to the small slepton masses, the amplitude is enhanced in 
smaller $E_l$ region, therefore 
the $A_l$ distribution is more flat than for the
$m_{\tilde{l}}=250$ GeV case. This suggests that $A_l$ 
and acceptance may depend on slepton masses when 
$\tchi^0_2\rightarrow \tilde{l}l$ is about to open.

It might be possible to determine the lepton energy asymmetry 
directly. In Ref. \cite{IK}, the correlation between the $A_l$ and 
$A_l^T$  distributions is studied for 
the $\tchi^0_2$ decay into $\tilde{l}_R l$.
Here 
$A_l^T\equiv |p^{l^+}_T-p^{l^-}_T|/(p^{l^+}_T+p^{l^-}_T)$, 
where $p^{l^+}_T$ and $p^{l^-}_T$ are transverse momenta of the 
two identified leptons. 
In the two body decay, $A_l$ in the
$\tchi^0_2$ rest frame is restricted to a rather 
small region. Although $\tchi^0_2$ originates 
from gluino or squark decays and should have 
some transverse momentum for the case 
studied in that paper, the $A^T_l$ distribution is still peaked 
where the $A_l$ distribution is located. This shows that $A^T_l$ 
reflects the decay distribution in the $\tchi^0_2$ rest frame. 

The acceptance would also depend on the $\tchi^0_2$ 
momentum distribution. If $\tchi^0_2$ has 
large transverse momentum, events with smaller 
$E^{\rm rest}_l$ would have a better chance of being accepted. 
Notice that if $E_l^{\rm rest}<p^{l{\rm cut}}_{T}$, 
this is the only possibility 
for the events being accepted; certainly the $\tchi^0_2$ 
transverse momentum distribution must be determined somehow 
to make a reliable estimate of the acceptance. 

Naturally, an average 
$p_{T\tchi^0_2}$ of the order of 
$m_{\tilde{g}(\tilde{q})}$ is expected when the $\tchi^0_2$ 
comes from $\tilde{q}$ or $\tilde{g}$ decay. 
The tree level kinematical distribution is determined  
by parent and daughter masses if the $\tchi^0_2$ 
originated  from a two 
body decay. Existing MC simulations describe 
the two body decay distributions correctly. 
For the three body decay, the distributions 
are different from those in phase space approximation. 
However, there would be no significant parameter 
dependence because only squark exchange 
contributes to the decay modes in Eq.~(\ref{e8}). 
If we have direct information on those masses, 
it may be rather easy to estimate the transverse momentum 
distribution\footnote{However, the transverse momentum of gluino 
receives (SUSY) QCD corrections. Correct estimation of 
the distribution would be very important.} 
of $\tchi^0_2$.

The $\tilde{\chi}^0_2$ 
distribution can be measured directly if there is
substantial statistics near the $m_{ll}$ end point. 
At the end point, the total momentum of the lepton pair is 
zero in the $\tchi^0_2$ rest frame, and we could reconstruct 
the velocity of $\tchi^0_2$, $\vec{\beta}_{\tchi^0_2}$, 
from the momentum of the lepton pair \cite{snowmass,hin1}. 
This $\tchi^0_2$ distribution 
may be convoluted with the $m_{ll}$ distribution and the $A_{l}$
distribution in the $\tchi^0_2$ rest frame to obtain the 
corresponding distributions in the laboratory frame. 
Even though the observed decay distributions 
are sensitive to the $\vec{p}_{\tchi^0_2}$ spectrum,
they might thus be corrected by measuring the
$\vec{\beta}_{\tchi^0_2}$ distribution using events with $m_{ll}$
near its end point.

So far we have only discussed the invariant mass distribution of 
lepton pairs from $\tchi^0_2$ decays, which is 
known to be important for SUSY studies at the LHC. In contrast, for the 
Tevatron upgrade, the trilepton mode is more important to 
discover supersymmetry \cite{tril, neut_taua}. This signal comes from the 
co-production of $\tchi^0_2\tchi^\pm_1$ and their decays into leptons. 
For this particular mode, the energy distribution 
of leptons from chargino decay must be considered in addition 
to the neutralino decay distributions.

\begin{figure}[htbp]
\begin{center}
\includegraphics[width=7cm,angle=-90]{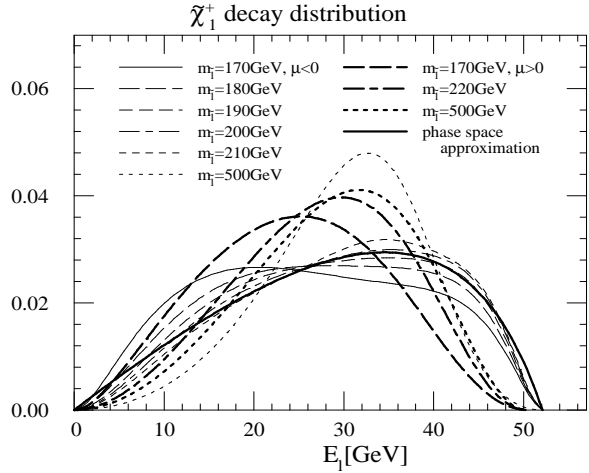}
\end{center}
\caption{\footnotesize
Energy distribution of the charged lepton from 
$\tchi^+_1\rightarrow\tchi^0_1 l^+\nu$ decay in the $\tchi^+_1$ 
rest frame. We take parameter set (A) 
($\mu<0$) and (B) ($\mu>0$) and vary $m_{\tilde{l}}$ 
as indicated in the figure. The thick solid line shows the distribution 
in the phase space approximation. The leptonic branching ratio 
of $\tchi^+_1$ is above 8\% for the parameters used in the figure.
}
\label{fig6}
\end{figure}

In Fig. 6, we show the lepton energy distribution 
of chargino $\tchi^+_1$ in the chargino rest frame
for the parameter sets (A) and (B). 
The amplitude of $\chi^+_1$ decay is easily obtained 
by replacing relevant masses and couplings in Eq. (\ref{e1}).
The parameter dependence is not very strong if the slepton mass 
is much above 200 GeV, but if it
is close to $\tchi^+_1$, the lepton energy distribution 
sensitively depends on the slepton mass. The branching 
ratio to the lepton is also larger in this region making the
trilepton mode more promising. For parameter 
set (A), the distribution is almost flat in $E_l$ for 
$m_{\tilde{l}}=170$ GeV. Such a dependence of the 
energy distribution in the $\tchi_1^\pm$ rest frame 
on MSSM parameters will affect 
the observed distribution of leptons in the lab frame. 
Notice that we must also pay attention 
to the effect of other cuts, e.g., on the total $E_T$ or
$\esla_T$. For example, for $\tchi^0_2\tchi^{\pm}_1$ 
co-production, the two $\tchi$ have balanced transverse momentum
unlike in the case where $\tchi^0_2$ comes from $\tilde{g}$ 
decay. 
The two $\tchi^0_1$ from $\tchi^0_2$ and $\tchi^\pm_1$ decays tend 
to be back to back to each other. This makes the total $\esla_T$ smaller, 
especially when the lepton energies are large. 
The correlation between various cuts must 
be studied very carefully.

In this section, we have only discussed the effect of  
lepton energy cuts on the observed $m_{ll}$ distribution, 
and its dependence on unknown MSSM parameters.
It is possible that other cuts like lepton isolation cuts, 
$\esla_T$ cuts, and total $E_T$ cuts correlate with
each other in a complicated manner to introduce additional
uncertainties. This may be corrected quite 
easily by existing MC simulations.  
The aim of this section has been to 
point out an obvious source of uncertainty that has not 
been taken into account in current MC simulations, 
and to propose several observables that might 
constrain this uncertainty directly. 

\section{Sensitivity to the Underlying Parameters}
In this section, we discuss if it is possible 
to extract the values of underlying MSSM parameters by measuring the 
$m_{ll}$ distribution. As we have already 
seen the distribution depends strongly on 
$m_{\tilde{l}}$, and also on the $\tchi^0_2\tchi^0_1Z$ and 
$\tchi_2^0 l\tilde{l}$ couplings. 
The latter is determined by the Higgsino components 
of $\tchi^0_2$ and $\tchi^0_1$. Because we take 
these neutralinos to be 
gaugino-like, their Higgsino components come from 
gaugino-Higgsino mixing. It depends on $\tan\beta$ and the 
Higgsino mass parameter $\mu$, and $|\mu|$ is roughly equal 
to $m_{\tchi^0_3}$, $m_{\tchi^0_4}$ or $m_{\tchi^+_2}$. Therefore 
the $m_{ll}$ distribution gives at least one 
constraint on $\mu$, $\tan\beta$ and $m_{\tilde{l}}$ in 
addition to the well known constraint on $m_{\tchi^0_2}-m_{\tchi^0_1}$. 

Because there is no MC simulation including the parameter 
dependence of three body decay distributions, 
we estimate the sensitivity 
to $\tan\beta$ and $m_{\tilde{l}}$ under the following 
{\it working} assumptions:
\begin{enumerate}
\item Backgrounds can be subtracted, or are negligible, and do not 
cause additional systematic errors.
\item The dependence of the acceptance on $m_{ll}$ can be corrected.
\end{enumerate}

Under these assumptions, we define the sensitivity 
function ${\cal S}$ as follows: 
\begin{eqnarray}\label{e9}
&&{\cal S}\left(M_1,M_2,
\mu, \tan\beta, m_{\tilde l_{L, R}}\vert_{\rm fit};
M_1,M_2,\mu, \tan\beta, m_{\tilde l_{L, R}}\vert_{\rm input}\right)
\nonumber \\
&&=\sqrt{\sum_i 
\left(n_i^{\rm fit}-n_i^{\rm input}\right)^2/n_i^{\rm input} }\, .
\end{eqnarray}
Here $n_i^{\rm fit}$ ($n_i^{\rm input}$) is the number 
of events in the $i$-th bin of the $m_{ll}$ distribution 
for the MSSM parameters 
$(M_1, M_2, \mu, \tan\beta, m_{\tilde l_{L, R}})
\vert_{\rm fit(input)}$. We normalize $\sum_i n_i^{\rm fit} 
= \sum_i n_i^{\rm input}$ to some number $N$.   
Then ${\cal S}$ gives the deviation of 
the input distribution $n_i^{\rm input}$  from the distribution 
for the fit ($n_i^{\rm fit}$) in units of standard deviations.
We take\footnote{
The available number of lepton pairs from $\tilde{g}$ decay 
ranges from $10^6$ to 1000 for 10 fb$^{-1}$ in the LHC study \cite{hin1}
for a grand unified theory (GUT) scale gaugino mass parameter $M=100$ GeV. 
At a high luminosity $e^+e^-$ collider, ${\cal O}(1000)$ lepton pairs 
could be produced. Our choice $N=2500$ should thus not be too
unrealistic.} 
$N=2500$ and an $m_{ll}$ bin size of 2 GeV.

\begin{figure}[htbp]
\begin{center}
\includegraphics[width=5.5cm,angle=-90]{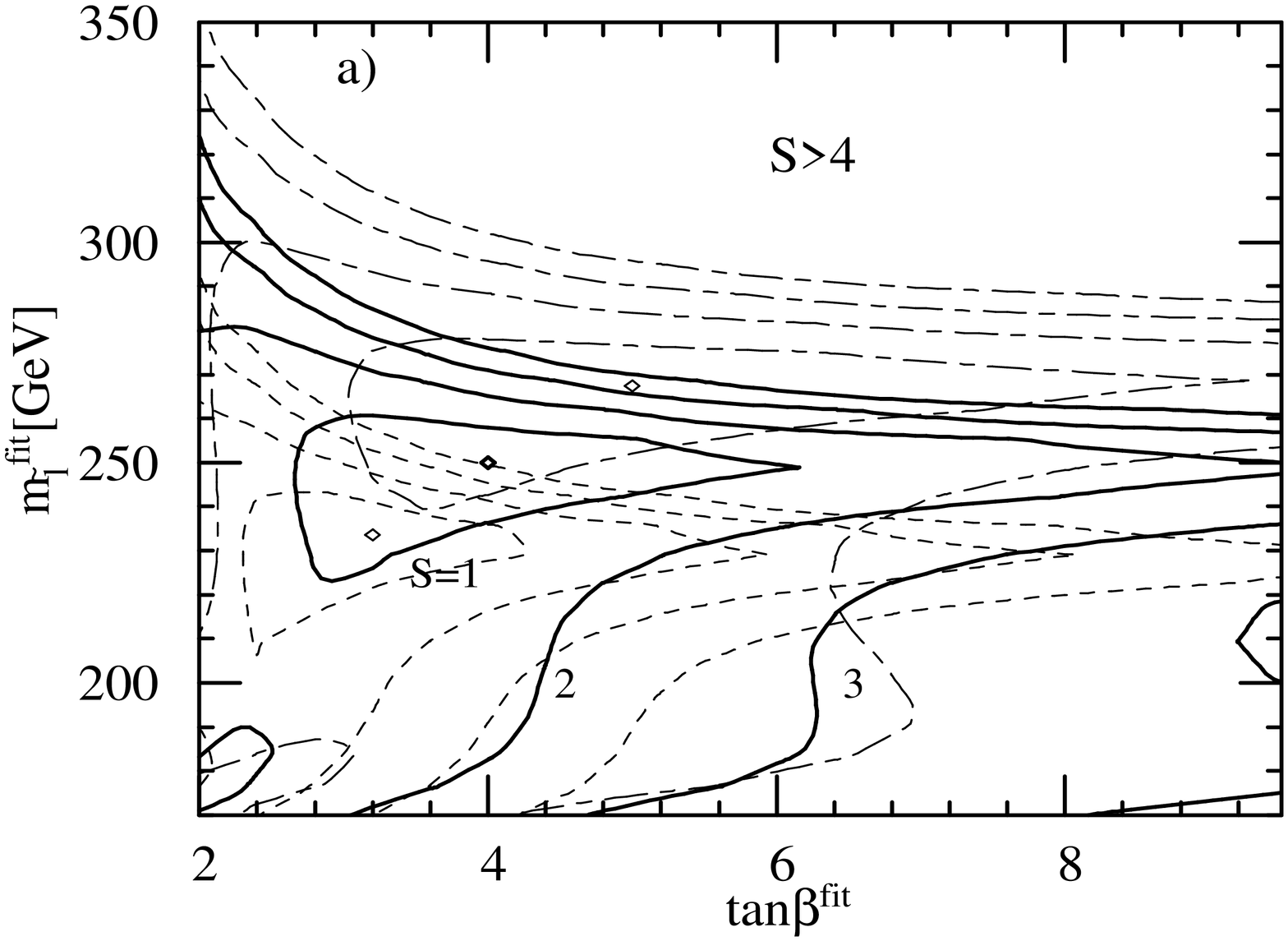}
\hskip 2cm 
\includegraphics[width=5.5cm,angle=-90]{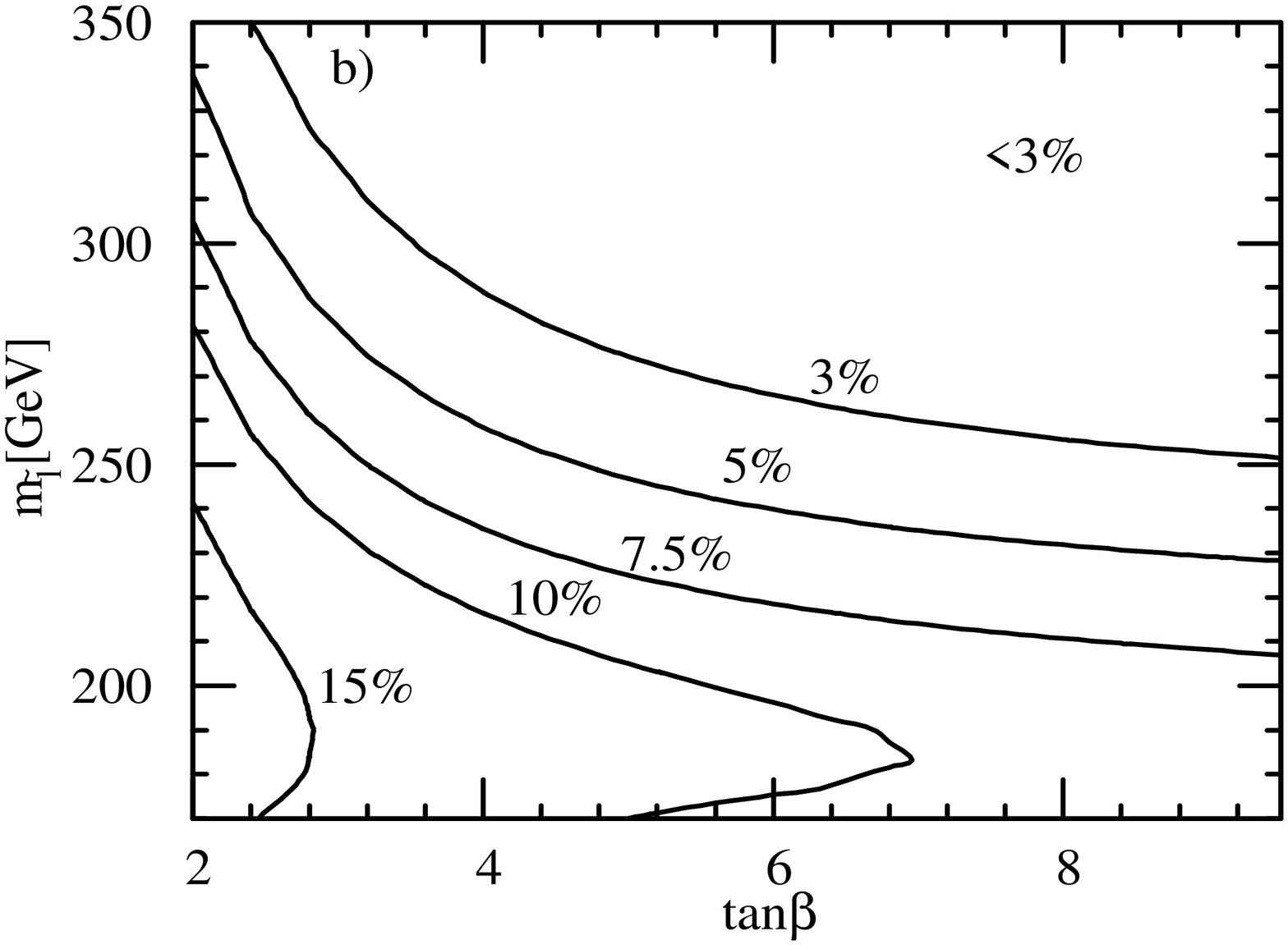}
\end{center}
\caption{\footnotesize 
(a) Contours of constant sensitivity function ${\cal S}=1,2,3,4$ 
in the ($\tan\beta^{\rm fit}$, $m_{\tilde{l}}^{\rm fit}$) plane. 
See text for the definition 
of ${\cal S}$ .
Input parameters are  set (A) with 
$m_{\tilde{l}}=250$ GeV. For solid lines, 
we fix $m_{\tchi^0_2}$, $m_{\tchi^0_1}$ and  $m_{\tchi^+_2}$ equal to 
those for parameter set (A) by varying $M_1^{\rm fit}$,
$M_2^{\rm fit}$ and   $\mu^{\rm fit}$, while 
$\tan\beta^{\rm fit}$ and $m_{\tilde{l}}^{\rm fit}$ are 
varied to see the sensitivity of the $\tchi^0_2$ 
decay distribution to these parameters.
For the dot-dashed (dashed) lines, $m_{\tchi^+_2}^{\rm fit}
=m_{\tchi^+_2}^{\rm input}+(-) 30$ GeV have been taken.
Diamonds correspond to the best fit points of each fit.
(b) Contours of constant $Br(\tchi^0_2\rightarrow
\tchi^0_1 l^+l^-)$ in the $\tan\beta$ and $m_{\tilde{l}}$ plane. 
The masses of $\tchi^0_2$, $\tchi^0_1$, and  $\tchi^+_2$ 
are set to equal to those for parameter set (A).
}
\label{fig7}
\end{figure}

In Fig. 7(a), we show contours of constant 
${\cal S}=1, 2, 3, 4$ (corresponding to 
1$\sigma$, 2$\sigma$, 3$\sigma$, 4$\sigma$) 
in the ($\tan\beta^{\rm fit}$, $m_{\tilde{l}}^{\rm fit }$) plane. 
For the solid lines, 
we take  parameter set (A) and $m_{\tilde{l}}=$ 
250 GeV as input parameters, while for fitting parameters 
we vary $\tan\beta$ and $m_{\tilde{l}_L}=m_{\tilde{l}_R}$, 
fixing ($M_1^{\rm fit}$, $M_2^{\rm fit}$,  
$\mu^{\rm fit}$) to reproduce the input values of
($m_{\tchi^0_1}$, $m_{\tchi^0_2}$, $m_{\tchi^+_2}$).\footnote{In 
our fit, we fix the ino masses, therefore 
we implicitly assume that the ``fake end point'' problem 
discussed in section 2 can be resolved when the
whole distributions are taken into account.}
The resulting contours (solid
lines) correspond to the sensitivity of the $m_{ll}$ distribution 
to $m_{\tilde{l}}$ and $\tan\beta$ when the three ino masses 
are known. 

In the figure, a strong upper bound on the slepton masses emerges,
while the lower bound 
is weak. Notice that there is another minimum near
$\tan\beta\sim 2$ and $m_{\tilde{l}}= 170$ GeV. 
By looking at the $A^T_{l}$ distribution, the two different 
minima may be distinguished. In that case a stronger lower bound would 
be obtained. On the other hand, 
$m_{\tilde{l}}<260 \ (270)$ GeV is obtained if ${\cal S}<1 \ (2)$ is
required. This can be understood as a result of the large change of the
distribution between $m_{\tilde{l}}=270$ GeV and $m_{\tilde{l}}=500$
GeV found in Fig. 1. 
The $m_{ll}$ distribution also constrains $\tan\beta$ mildly. 
The constraint is not very strong due to our 
choice of parameters $|\mu|\gg M_2$; gaugino-Higgsino  
mixing is suppressed in this case.

We computed these contours by fixing the three ino masses. This may not be a
realistic assumption, because $\tchi^\pm_2$  
might not be observed at the LHC. If we perform the fitting with 
varied $m_{\tchi^+_2}^{\rm fit}$, 
we can find almost the same $m_{ll}$ distribution 
for different sets of parameters. In Fig. 7, the dashed (dot-dashed) 
lines correspond to constant ${\cal S}$ 
contours when $m_{\tchi^+_2}^{\rm fit}= m_{\tchi^+_2}^{\rm input}-30$
GeV ($m_{\tchi^+_2}^{\rm fit}= m_{\tchi^+_2}^{\rm input}+30$ GeV) 
is required for the 
fitting. The best fit points are $\tan\beta =3.2$ and 
$m_{\tilde{l}}=233.6$ GeV for 
$m_{\tchi^+_2}^{\rm fit}= m_{\tchi^+_2}^{\rm input}-30$ GeV,
and $\tan\beta= 5$ and $m_{\tilde{l}}=267$ GeV for
$m_{\tchi^+_2}^{\rm fit}= m_{\tchi^+_2}^{\rm input}+30$ GeV.  
The best fit point therefore 
moves from smaller $\tan\beta$ and $m_{\tilde{l}}$  
to larger ones as $m_{\tchi^+_2}^{\rm fit}$ is increased. 
Other distributions such as $p^l_{T}$ and $A^T_{l}$
might differ at different best fit points; however, we do not go 
into the details of these distributions. 

At each best fit point with different $m_{\tchi^+_2}^{\rm fit}$,
${\cal S}\sim 0$. 
Therefore the dashed or dot-dashed  contours 
are very similar to the contours of constant ${\cal S}$ 
where the parameters for the best fit point are taken as the inputs. 
It can be seen that 
as $m_{\tchi^+_2}$ is decreased, the constraint on $\tan\beta$ becomes 
stronger, because Higgsino-gaugino mixing becomes more important. 
On the other hand, when $m_{\tchi^+_2}^{\rm fit}$ is increased, 
the constraint on $\tan\beta$ tends to disappear.

The constraint from the $m_{ll}$ distribution is independent of 
that from the decay branching ratio. In Fig. 7(b), we 
show the contours of constant 
$Br(\tchi^0_2\rightarrow\tchi^0_1l^+l^-)$. 
Notice that the branching ratio does not 
bound $\tan\beta$ by itself unless 
$Br(\tchi^0_2\rightarrow\tchi^0_1l^+l^-)> 10$\%.
The shapes of these contours are not particularly correlated 
to the constraint from the $m_{ll}$ distribution 
shown in Fig. 7(a); therefore combining the two pieces of
information might constrain the parameter space 
further. However, we should be aware that the branching 
ratio depends on the whole sfermion mass spectrum. 
For example, if there is a substantial 
reduction of masses for third generation 
sfermions, the branching ratio will be changed. 
Fits using the $m_{ll}$ distribution are very important 
in that sense, because the shape depends 
only on ino mass parameters and $m_{\tilde{l}}$. 
Notice also that  we only measure the sum of 
the products of several branching ratios rather than 
$Br(\tchi^0_2\rightarrow \tchi^0_1 l^+l^-)$ itself at the LHC. 

\section{Discussion and Conclusion} 
In this paper, we examined the impact of neutralino 
decay distributions on the study of the minimal supersymmetric 
standard model at the CERN LHC. The leptonic three body decay 
of the second lightest neutralino, 
$\tchi^0_2\rightarrow l^+l^- \tchi^0_1$, is known to be very 
important because the end point of the lepton invariant 
mass distribution $m_{ll}^{\rm max}$ gives us direct 
information about the mass difference between $\tchi^0_2$ 
and $\tchi^0_1$, which gives us a stringent constraint 
on MSSM parameters. 

We found that the neutralino decay distribution 
depends on the slepton masses rather sensitively. 
Measuring the shape of the lepton invariant mass distribution 
can be important even for the determination of $m_{ll}^{\rm max}$. In
some cases, the measured end point may not coincide with 
the neutralino mass difference $m_{\tchi^0_2}-m_{\tchi^0_1}$,
due to the strong suppression of the amplitude 
near $m_{ll}^{\rm max}$, while the leptonic branching 
ratio is still around a few \%.

On the other hand, if systematic errors can be removed, 
the distribution gives us a model independent probe of the slepton 
mass scale. For the $\tchi^0_2$ signal 
from $\tilde{g}$ decay, the $S/N$ ratio is very large, 
or the background can be subtracted by using
lepton pairs with opposite charge and different flavors. 
The remaining distribution can be further studied by looking at 
(A) the $\vec{\beta}_{\tchi^0_2}$ distribution 
measured by using events which have $m_{ll}$ near its end point, and 
(B) the lepton transverse momentum asymmetry $A^T_{l}$ 
distribution, which is well correlated with the lepton energy 
asymmetry in the $\tchi^0_2$ rest frame. 

Notice that if the $\tchi^0_2$ momentum distribution 
is precisely measured, leptonic decay distributions 
may be discussed without any QCD uncertainty, which might otherwise be 
substantial. Of course, the
correlation with other cuts (e.g., on $\esla_T$,
$\sum {E_T}$, lepton isolation) must be either small or 
determined from direct measurements, so that 
measurements in the lab frame allow us to reconstruct 
the $\tchi^0_2$ decay distribution in its rest frame.  
The revenue of such an effort to reduce the uncertainty from 
the cuts is information on slepton 
masses and neutralino mixings, independent of any assumption
about the mechanism to break supersymmetry.

In order to see if a study in such a direction is possible, dedicated MC 
simulations are necessary. Notice that the commonly available 
MC codes ISAJET and SPYTHIA used to simulate the three body decay 
distribution in phase space approximation. Our results show that a
more careful treatment of $\tchi^0_2\rightarrow 
l^+l^-\tchi^0_1$ decays is important, if one is interested in 
decay distributions. Moreover,
the acceptance of such di-lepton events depends 
on the lepton energy in $\tchi^0_2\rightarrow l^+l^-\tchi^0_1$ 
decays.

The study of $\tchi^0_2$ decays can be easily done at 
$e^+e^-$ colliders. $\tchi^0_2\tchi^0_2$ pair production 
or $\tchi^0_2\tchi^0_1$ co-production do not 
suffer from large backgrounds. Though the statistics 
is rather limited there, it would give us 
constraints on slepton masses and neutralino mixings. 
Notice that in supergravity models
the lighter chargino and neutralinos 
are expected to be lighter than squarks and gluinos. 
Both LC and LHC may find and study $\tchi^0_2$, 
and the information gleaned from these analyses
can be combined to obtain 
a better understanding of MSSM parameters and the
SUSY breaking mechanism.

\section*{Acknowledgements}
We thank S. Kiyoura for earlier collaboration and discussion, 
H. Baer on the information about most recent ISAJET status, and 
M. Drees and X. Tata for comments and careful reading of manuscript. 
This work was supported in part by the Grant-in-Aid for Scientific 
Research (10140211 and 10740106) from the Ministry of Education, 
Science, Sports, and Culture of Japan.
Y.Y. was also financially supported in part by Fuuju-kai Foundation. 

\section*{Appendix}
\setcounter{equation}{0}
\renewcommand{\theequation}{A\arabic{equation}}

We show the explicit forms of the neutralino couplings used 
in section 2. 

The neutralino mass matrix $M_N$ in the gauge eigenbasis 
$(\tilde{B},\widetilde{W}^3,\tilde{H}^0_1,\tilde{H}^0_2)$ is 
written as 
\begin{equation}\label{e10}
M_N= \left(
\begin{array}{cccc}
M_1& 0& -m_Zc_{\beta}s_W & m_Zs_{\beta}s_W\\
0& M_2& m_Z c_{\beta}c_W & -m_Z s_{\beta}c_W\\
-m_Z c_{\beta}s_W & m_Z c_{\beta}c_W & 0 & -\mu\\
m_Z s_{\beta}s_W & -m_Z s_{\beta}c_W & -\mu & 0         
\end{array}
\right)\ ,
\end{equation}
where $s_W=\sin\theta_W$, $c_W=\cos\theta_W$, $s_{\beta}=\sin\beta$, 
and $c_{\beta}=\cos\beta$.
The mass eigenstates $\tchi^0_i$ ($i=1-4$) are related to the 
gauge eigenstates via the mixing matrix $N_{ij}$ as 
\begin{equation}\label{e11}
\tchi^0_i= N_{i1}\tilde{B}+N_{i2}\widetilde{W}^3+N_{i3}\tilde{H}^0_1
+N_{i4}\tilde{H}^0_2.
\end{equation}
We neglect the possibility of
CP violation for sparticles and take $N_{ij}$ as a real matrix 
by allowing negative values of $m_{\tchi^0}$ in Eq.~(\ref{e4}). 

The interaction Lagrangian of fermion-sfermion-neutralino 
and neutralino-$Z^0$ couplings is written as
\begin{equation}\label{e12}
  {\cal L}_{\rm int}
%=  -g_2 \overline{\tilde \chi^0_A} (N^{R(f)}_{AX} P_R +N^{L(f)}_{AX} P_L) f
%    \tilde f_X^\dagger + ({\rm H.c.}),
=  -g_2 \overline{\tilde \chi^0_A} (a^f_{AX}P_L + b^f_{AX}P_R) f
    \tilde f_X^\dagger + ({\rm H.c.})
   +\frac{g_Z}{2}z^{(\tchi^0)}_{BA}\overline{\tchi^0_B}
    \gamma^{\mu}\gamma_5\tchi^0_A Z^0_{\mu},
\end{equation}
where $f$ stands for $l$, $\nu$, $d$, and $u$. 
The couplings ($a$, $b$) for leptons are, 
in the ($\tilde{l}_L$, $\tilde{l}_R$) basis, 
\begin{eqnarray}
%  N^{L(l)}_{AX}&=& \frac{1}{\sqrt{2}} \{
  a^l_{AX}&=& \frac{1}{\sqrt{2}} \left\{
       [-N_{A2} -N_{A1} t_W] \delta_{X,L}
        + \frac{m_{l}}{m_Wc_\beta} N_{A3} \delta_{X,R} \right\},
\nonumber \\
%  N^{R(l)}_{AX} &=& \frac{1}{\sqrt{2}} \{ 
  b^l_{AX} &=& \frac{1}{\sqrt{2}} \left\{ 
           \frac{m_{l}}{m_W c_\beta} N_{A3} \delta_{X,L} 
           +2 N_{A1} t_W \delta_{X,R} \right\},
\nonumber \\
  a^{\nu}_{AX} &=& \frac{1}{\sqrt{2}}
             [N_{A2}-N_{A1} t_W] \delta_{X,L}, 
\nonumber \\
  b^{\nu}_{AX}&=&0, \label{e13}
\end{eqnarray}
where $t_W=\tan\theta_W$. 
We have always ignored ${\cal O}(m_l)$ terms in Eq.~(\ref{e13}) 
in our numerical calculations.

The coupling of $\tchi^0$ and $Z^0$ takes the form 
\begin{equation}\label{e14}
z^{(\tchi^0)}_{BA} = N_{B3}N_{A3}-N_{B4}N_{A4}. 
\end{equation}

\end{document}